\documentclass[12pt]{article}

\usepackage{url}

\usepackage{graphicx}
\usepackage{authblk}
\usepackage[margin=1in]{geometry}
\usepackage{cite}

\begin{document}

\title{Lying particles}


\author[1]{Lev Vaidman}
\affil[1]{\small Raymond and Beverly Sackler School of Physics and Astronomy, Tel-Aviv University, Tel-Aviv 69978, Israel}

\date{}

\maketitle

\begin{abstract}
The common feature of several experiments, performed and proposed, in which particles provide misleading evidence about where they have been, is identified and discussed. It is argued that the experimental results provide a consistent picture when interference amplification effects are taken into account.
\end{abstract}


\section{Introduction}

Perhaps the most significant difference between classical and quantum physics is that quantum particles do not have well-defined trajectories (unless we accept an arguably fringe Bohmian interpretation, \cite{Bohm}). Standard quantum mechanics does not have a clear concept of the location of a particle. There is no answer to the question: Through which slit did the particle pass in a two-slit interference experiment? Still, there is a growing interest in discussing which-path questions for particles passing through interferometers, notably following Wheeler's controversial proposal, \cite{Wheeler}.

Wheeler suggested assigning a well-defined path to the particle in the interference experiment of the wave packets when only one wave packet had a continuous trajectory from the source to the detector.
I proposed an alternative approach, \cite{past}, in which the particle was present in the locations where it left a significant trace in the environment. According to my definition, a pre and postselected particle can be in several places simultaneously, and, moreover, the particle might have disconnected regions of presence. My proposal created a significant controversy, \cite{Li13,RepLiCom,Grif16,RepGrif,SalihCH,Hash,HashCom,HashComRep,Eli,Dupr,Disapp,ACWE,
Hance23,HanceCom,HanceComRep}.

To demonstrate the usefulness of my definition,
I led an experiment, \cite{Danan}, in which the photons themselves told us where they had been inside an interferometer, which only increased the controversy, \cite{Berge,BergeCom,BergeComRep,AskingphotonKedem,Hasegawa,myphotonsneutrons,Bart,BartCom,Poto,PotoCom,China,Ben-Israel2017,Sok,SokCom,Sali,SaliCom}. Although the experiment was supposed to demonstrate the local trace left by photons in different locations, because of the difficulty of performing conditional counting, the measurement pointer was the photon itself: the photon's transversal degree affected by the local interaction with the environment. The justification of this modification in the experiment of \cite{Danan} was the fact that the record in this degree of freedom of the photon was not distorted during the time the photon wave packet was moving toward the detector.

I also argued that there was another way to interpret this experiment as a demonstration of the presence of the particle in various interferometer paths.  All interactions are local. If the pre and postselected photons bring information about a disturbance which was solely introduced in one location, then the photon tells us that it was there. This sounds like a very strong argument, but it turned out to be a subtle issue. This is what I analyze in this report. The analysis demonstrates the core of the resolution of the debates \cite{Nik} and \cite{Nikrep}, \cite{Jordan} and \cite{JordanCom}, \cite{Bhati} and \cite{BhatiCom}, \cite{Yuan} and \cite{YuanCom}.

\section{The general setup}

We consider a particle entering an interferometer with multiple paths in a particular input state that reaches a particular output port. The interferometer has mirrors and beam splitters that cause a well-defined unitary evolution of the degree of freedom of the path of the particle. In addition, the particle has another degree of freedom with a specified state $|\Phi \rangle$ at the input port which is observed at the output port. For my analysis, I assume that if the interferometer is undisturbed, the state of this degree of freedom remains unchanged everywhere inside the interferometer. The presence of the pre and postselected particle in a particular path of the interferometer is then characterized by the effect of the local disturbance in this path on the state of the particle at the output port.  Any disturbance of the non-path degree of freedom in a particular path can always be expressed as
\begin{equation}\label{ref}
|\Phi\rangle\rightarrow\mathcal{N}(|\Phi\rangle +   \epsilon |\Phi_{\perp}\rangle),~~~~\epsilon \ll 1,
\end{equation}
 where $\langle\Phi|\Phi_{\perp}\rangle=0$. The non-path degree of freedom of the state in the output port is:
\begin{equation}\label{definition}
|\Phi\rangle_{out}=\mathcal{N}(|\Phi\rangle + \alpha  \epsilon |\Phi_{\perp}\rangle).
\end{equation}
The parameter $\alpha$ characterizes
the modification of the effect of the interaction and thus the amount of presence of the particle in the respective path.

\begin{figure}
\centerline{\includegraphics[width=0.9\textwidth]{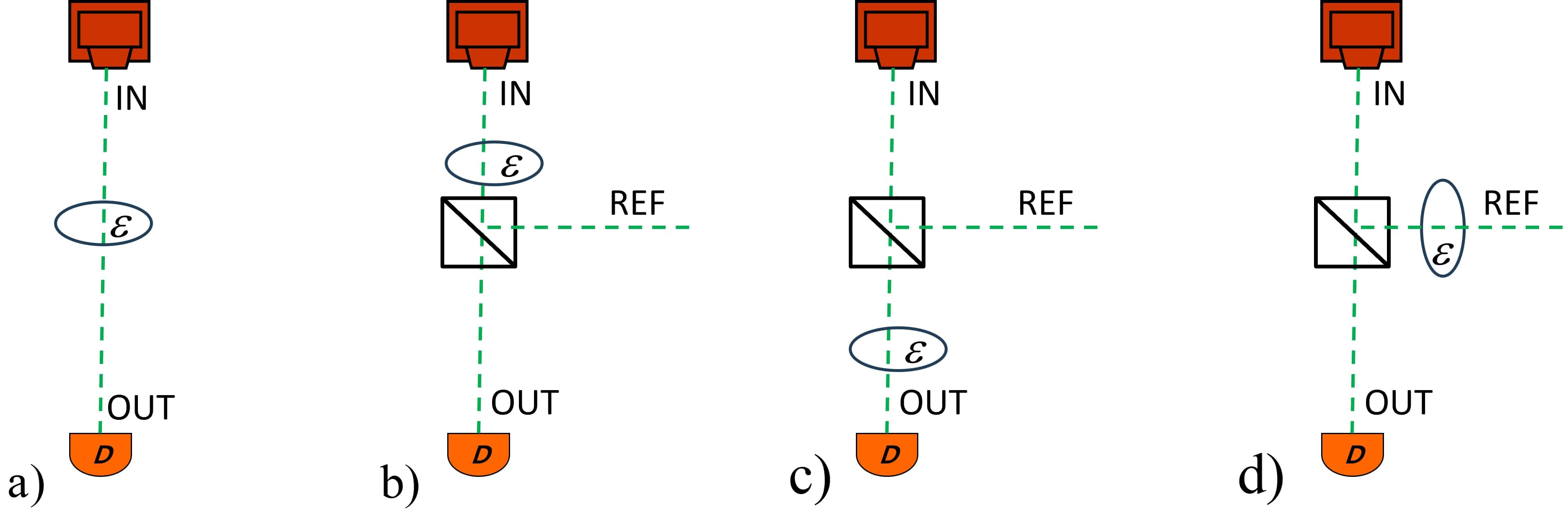}}
\caption{ {\bf A particle tells us about its presence.}
a) Reference signal from the path certainly taken by the particle. b)-d) Obtaining the signal of presence of the particle (modification of the state reaching the detector) in different paths of pre and postselected particle when a beam splitter is added. The ellipse with $\epsilon$ indicates the location where the disturbance is introduced to test for the presence of the particle.
}
\end{figure}

As expected, the one-path ``interfereometer'' has a measure of presence in the path $\alpha=1$, Fig. 1a.
 If we add a beamsplitter, Fig. 1b-d, but consider the particle which reaches the detector, then the particle has  presence
\begin{equation}\label{1bd}
\alpha_{\rm IN}=\alpha_{\rm OUT}=1,~~ \alpha_{\rm REF}=0.
\end{equation}

  If we consider a balanced Mach-Zehnder interferometer (MZI) Fig.~2a, tuned to constructive interference, then we are not surprised to obtain \begin{equation}\label{2a}
\alpha_{\rm IN}=\alpha_{\rm OUT}=1,~~ \alpha_A=\alpha_B=\frac{1}{2}.
\end{equation}

 If MZI has beam splitters reflecting $90\%$, which is also tuned to constructive interference, Fig. 2b, we obtain \begin{equation}\label{2b}
\alpha_{\rm IN}=\alpha_{\rm OUT}=1,~~ \alpha_A=0.9,~~\alpha_B=0.1.
\end{equation}

  If we tune this interferometer to the minimum intensity at the detector, Fig. 2c, then the experiment shows marks of the presence in the arms of the interferometer which are less intuitive
    \begin{equation}\label{2c}
\alpha_{\rm IN}=\alpha_{\rm OUT}=1,~~ \alpha_A=\frac{9}{8},~~\alpha_B=-\frac{1}{8}.
\end{equation}

However, I do not consider these results to be incorrect. They provide the same characterizations of the presence of a pre and postselected particle at particular locations as given by the weak value of the projection on these locations. They describe the universal property of modifications of all weak couplings at these locations, \cite{PNAS}.  The particles do not lie about their presence (in the above sense), which can be larger than 1 or negative, or even corresponding to a complex number. The presence can be anomalously large; the only constraint is that at any moment of time one particle should have a total presence in all paths of an interferometer equal to 1.

\begin{figure}
\centerline{\includegraphics[width=0.9\textwidth]{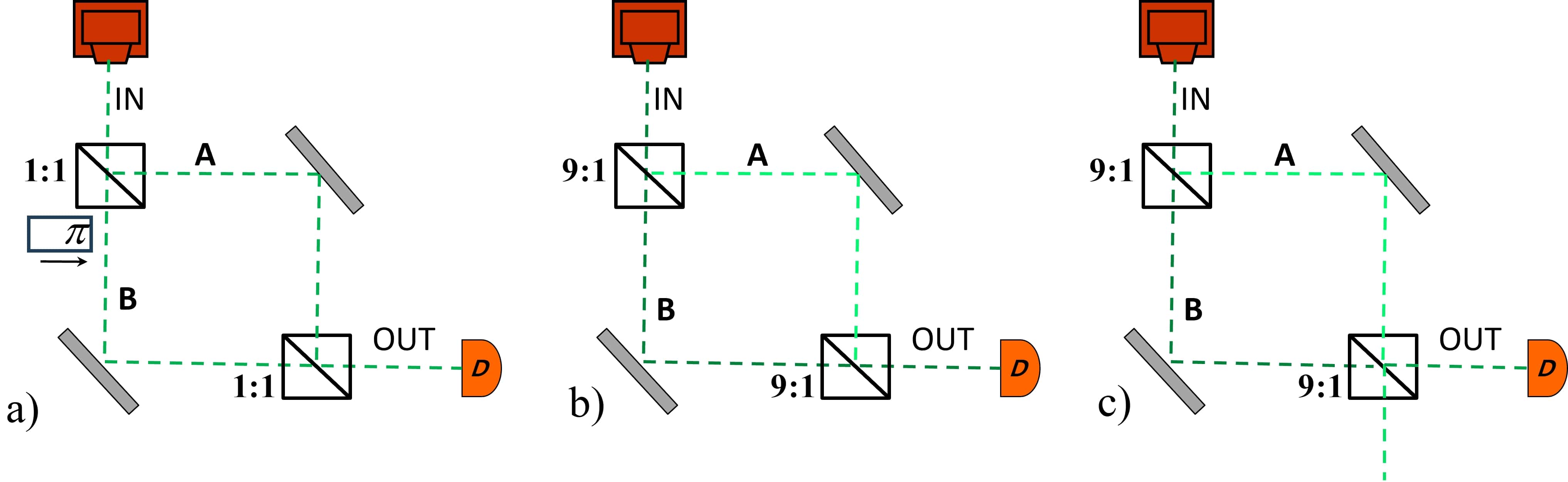}}
\caption{ {\bf A  Mach-Zehnder interferometer.}
a) Balanced MZI tuned to constructive interference.  [A device inserted into arm $B$ which introduces relative phase $\pi$ in the non-path degree of freedom makes the particles lie about its presence.] b) Unbalanced MZI tuned to constructive interference. c) Unbalanced MZI tuned to maximally destructive interference.}
\label{fig:2}
\end{figure}

  Particles also do not lie when they tell us about a surprising picture of presence in nested MZI, \cite{past,Danan}, see Fig.~3a.
  \begin{equation}\label{3a}
 \alpha_{\rm IN}=\alpha_{\rm OUT}=1,~~\alpha_C =1,~~ \alpha_E =\alpha_F=0,~~ \alpha_A =1,~~\alpha_B=-1.
\end{equation}
I consider these results to be a correct demonstration of the presence of the particle in the interferometer, since it faithfully represents the trace left on the paths by the particle. Every mirror in the path gets a kick equal to $\alpha\sqrt{2} p$ where $p$ is the momentum of the particle. (Note that in experiment \cite{Danan} a special arrangement with disturbances with different frequencies was implemented. This allowed the measurement of the presence in all paths simultaneously, but instead of $\alpha$, only the absolute value of the signal $|\alpha|$ was observed.)

\begin{figure}[t]
\centerline{\includegraphics[width=0.9\textwidth]{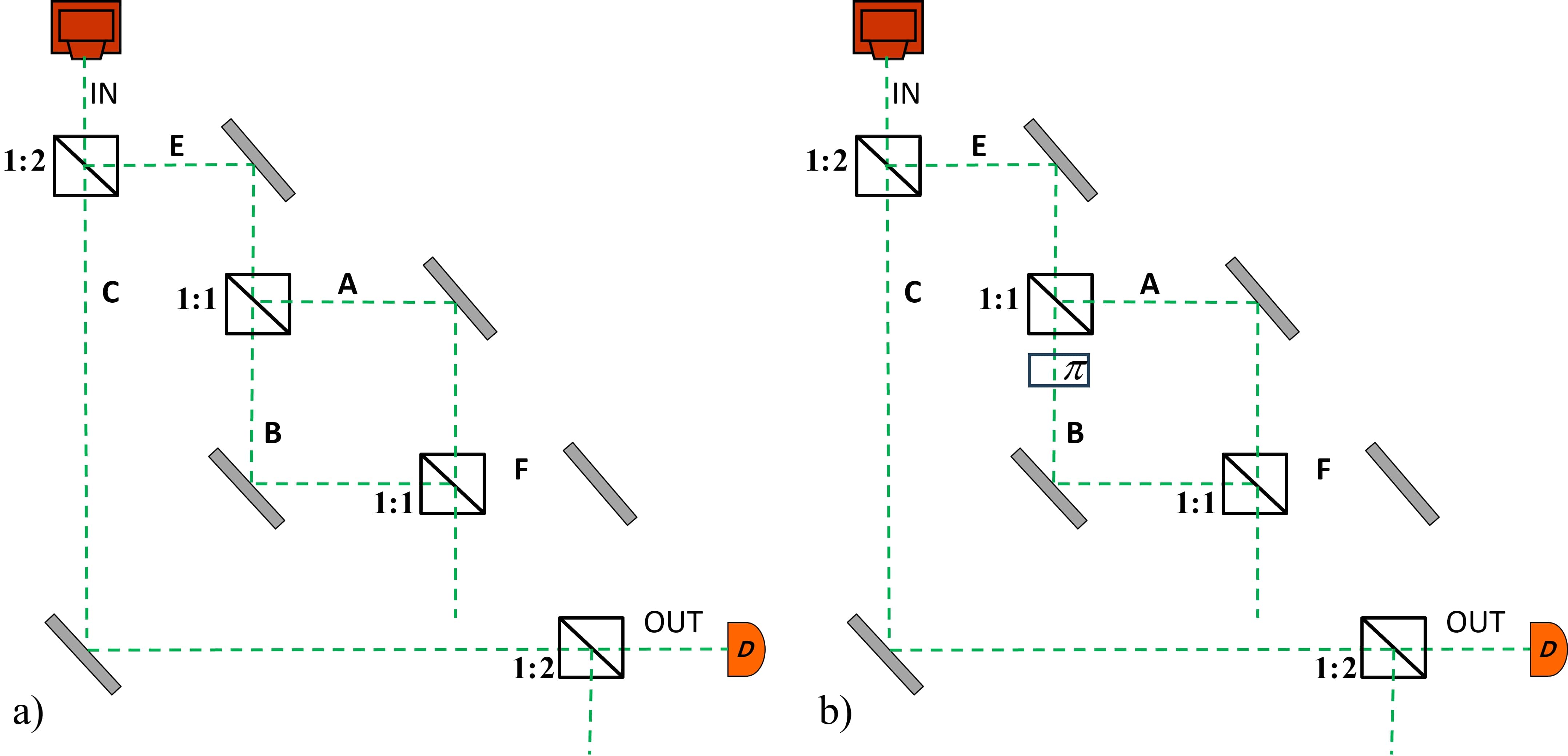}}
\caption{ {\bf Nested  Mach-Zehnder interferometer.}
a) The interferometer with a particle present in a disconnected region, cf. Fig.~2b. of \cite{Danan}. b) The same interferometer with a device introduced in path E which affects the non-path degree of freedom of the particle making it to lie about where it has been. The traces left by the particle on the parts of the interferometer are essentially the same in the two cases.}
\label{fig:3}
\end{figure}

\section{Interferometers with particles lying about where they have been}

Let us now consider cases where particles lie about their presence in a particular path of the interferometer. Our setup is the same as above, but we add in one of the paths a device which affects the non-path degree of freedom in the following way. It adds the relative phase $\pi$ to the component $|\Phi_{\perp}\rangle$ which is created by the disturbance (\ref{ref}):
\begin{equation}\label{pi}
a|\Phi\rangle +b    |\Phi_{\perp}\rangle\rightarrow a|\Phi\rangle -b    |\Phi_{\perp}\rangle.
\end{equation}
This corresponds to a modification of the experiment of \cite{Danan} proposed by \cite{Jordan}. They inserted a Dove prism inside an interferometer. The non-path degree of freedom of the particle was its transversal motion of the photon. The disturbance was a transversal shift. The shift of a well-aligned Gaussian profile created an orthogonal component which was an odd function in the transverse direction. The action of the Dove prism on such a component is to flip its sign, i.e., to add phase $\pi$. The other proposals (\cite{Nik,Bhati,Yuan}) have an equivalent mathematical structure.

Introducing the relative phase to the component that is not present changes nothing, so placing the device makes no effect if the interferometer is not disturbed, but it causes the particles to lie (in some cases) about where they have been by spoiling the mathematical equivalence between the local trace on the environment and the observed trace on the non-path degree of freedom of the particle. The records created locally on the particle are distorted during the passage of the particle to the output port. I will show two examples of how it happens.

Let us add this device at the beginning of the path $B$ of the balanced MZI described in Fig.~2a.
 Then, our method of characterization of the presence of the particle by observing the effect of introducing the disturbance (\ref{ref}) in different paths on the non-path degree of freedom at the output port will lead, instead of (\ref{2a}) to
   \begin{equation}\label{2a'}
\alpha_{\rm IN}=0,~~ \alpha_{\rm OUT}=1,~~ \alpha_A=\frac{1}{2},~~\alpha_B=-\frac{1}{2}.
\end{equation}
Clearly, this description does not make sense. We know that the particle entered the interferometer. The method itself tells us that it left the interferometer, but the method also tells us that the particle was not present at the input port.
 The malfunction of the method is transparent: the orthogonal component created at the input port splits into a superposition in the arms $A$ and $B$ and the phase device in the arm $B$ causes destructive interference in the output port.
The absence of the signal from the input port does not ensure that the particle was not there.

The second example is a nested interferometer of \cite{Danan}, which has already been the subject of great controversy. Fig.~3b shows explicitly the phase $\pi$ device added in the arm $B$.  Then, our method of characterization of presence by observing the effect  leads instead of (\ref{3a}) to
\begin{equation}\label{3b}
\alpha_{\rm IN}=3, \alpha_{\rm OUT}=1,~~ \alpha_C =1,~~ \alpha_E =2,~~\alpha_F=0,~~ \alpha_A =1,~~\alpha_B=-1.
\end{equation}
These results are neither present a sensible classical picture, nor describe faithfully a modification of the weak couplings of the particle with the environment which are, in fact, not affected by the phase shifting device and remain to be described correctly by (\ref{3a}).

\section{Discussion}

In Fig.~3b. I show the experiment with a nested interferometer in which we get a signal from the path $E$, but I claim that it was not there. How can the particle bring a signal from the place it was not present? It contradicts the idea of causality and locality of all interactions.

The answer to this question is not simple. In fact, there is some kind of particle presence in $E$, also in the interferometer described in Fig.~3a without $\pi$ shifter, which I named ``secondary'' presence, \cite{morepast}. The main property of the secondary presence is that although there is no trace of the first order (in $\epsilon$) in $E$, placing a block in $E$ will affect the first order traces elsewhere (in $A$ and $B$). Also, when the interferometer is not ideal (and there are no ideal interferometers) there will be a higher order trace in $E$.

The question to be answered is how the phase shifter causes particles to lie about their presence showing a strong signal in $E$ that looks like a signal of ordinary presence. This is due to the anomalous sensitivity to the presence in $E$ when the phase device is inserted. If we add a nondemolition measurement of the presence in $E$ and condition our observation of the signal on finding the particle in $E$, our signal instead of $\alpha_E =2$ will be $\alpha_{E~[{\rm conditioned~on~presence~in~E}]} \rightarrow \infty$. This follows from the definition of $\alpha$ in (\ref{definition}), since the state at the output port is $|\Phi_{\perp}\rangle$.

The locality of interaction tells us that if the particle brings a signal from a particular location, the particle must have some presence there, but the ``amount of presence'' is not characterized directly by this signal. It is characterized by the ratio between the obtained signal and the reference signal that would be obtained if the particle were localized there. Since in the nested interferometer with the phase shifter the signal is finite, but the localized particle in this experiment leads to an unbounded signal, the ratio is zero. The method tells us that the particle was not present there. See the analysis of a realistic experiment in \cite{Yuan,YuanCom}.

I suggest adopting the definition according to which the particle passing through an interferometer was in a particular path if a local disturbance in this path (if it was performed) can be observed in the output state of the
particle. Every path which fulfills this criterion is the path where, by this definition, the particle was. In many cases, then, the particle is present in more than one path simultaneously. This definition contradicts the fact that a single particle is never found in two places simultaneously.
The setup I proposed cannot demonstrate this contradiction, because the proposal introduces disturbances in the paths one by one and not simultaneously together. Let us see what happens if we introduce disturbances in two places simultaneously.

We consider a particle passing MZI described in Fig.~2a and try to observe its presence in arms $A$ and $B$. To obtain evidence that it is present simultaneously
in two places we need that the particle will have two non-path degrees of freedom $|\Phi_A\rangle$ and $|\Phi_B\rangle$ affected by the disturbances in the two paths. Then, the non-path degree of freedom state of the particle will be
\begin{equation}\label{finalAB}
|\Phi_A\rangle|\Phi_B\rangle \rightarrow \mathcal{N}(|\Phi_A\rangle|\Phi_B\rangle  + \frac{\epsilon}{2} |\Phi_{A\perp}\rangle|\Phi_B\rangle + \frac{\epsilon}{2} |\Phi_{A}\rangle|\Phi_{B\perp}\rangle).
\end{equation}

There is a first order signal both from path $A$ and from path $B$, so at this stage, by definition, the particle was in two places. However, we will never be able to find one particular particle in two places simultaneously.
Performing analysis of this state on an ensemble of particles starting at the source and ending up at the output port we will observe these first order traces, so we will be able to use the definition. This, however, does not provide direct information about every particle in the ensemble. A particular measurement which occasionally (with probability larger than 0) finds state $|\Phi_{A\perp}\rangle$ tells us that the particle was in path $A$ and finding $|\Phi_{B\perp}\rangle$ tells us that the particle was in path $B$. The corresponding degrees of freedom in state (\ref{finalAB}) are entangled and the term $|\Phi_{A\perp}\rangle|\Phi_{B\perp}\rangle$
is not present.

We do not have to adopt the many-world interpretation \cite{SEP} to appreciate the result of this paper, but it seems to me that the MWI provides the most satisfactory resolution of the apparent paradox of claiming that the particle in some sense is in two places simultaneously, in spite of the fact that we are not able to find the particle simultaneously in two places.
In a world the particle passes through the interferometer, it has ``memory'' of order $\epsilon$ of being both in $A$ and in $B$. However, at the moment when a measurement verifying the presence of $|\Phi_{A\perp}\rangle$ is performed, the world splits into two, one in which we know that it was in $A$ and one in which we have no decisive information. Due to the entanglement of degrees of freedom, in the world we know about the presence in the path $A$ the component
 of order  $\epsilon$
signifying the presence in path $B$ is erased. This is why we never find definite evidence of simultaneous presence in the two arms of the interferometer.

\section*{Acknowledgments}
I thank Jan Dziewior and Gregory Reznik
for very helpful discussions. This work has been supported in part by the National Science Foundation Grant No. 1915015 and the U.S.-Israel Binational Science Foundation Grant No.~735/18 and by the Israel Science Foundation Grant No.~2064/19.




\bibliographystyle{unsrt}

\bibliography{arxiv}


\end{document}